\documentclass[letterpaper,superscriptaddress,english,aps,prx,twocolumn, amsfonts, amssymb]{revtex4}
\usepackage{epsfig, graphicx,graphics,amsmath,amssymb,float}
\usepackage[T1]{fontenc}
\usepackage[latin9]{inputenc}
\usepackage{amsmath}
\usepackage{amssymb}
\usepackage{amscd}
\usepackage{bm}
\usepackage{psfrag}
\usepackage{times}
\usepackage{bbm} 
\usepackage[caption=false]{subfig}
\usepackage{subfig}
\usepackage{color}
\usepackage[normalem]{ulem}
\usepackage[bookmarks=true,colorlinks,linkcolor=blue,urlcolor=blue,citecolor=blue]{hyperref}

\newcommand{\be}{\begin{equation}}
\newcommand{\ee}{\end{equation}}
\newcommand{\bea}{\begin{eqnarray}}
\newcommand{\eea}{\end{eqnarray}}
\newcommand{\ket}{\rangle}
\newcommand{\bra}{\langle}

\definecolor{darkpink}{rgb}{0.91, 0.33, 0.5}

\begin{document}
\title{Confinement and bound states of bound states in a transverse-field two-leg Ising ladder}
\author{Fl\'avia B. Ramos}
\affiliation{International Institute of Physics,  Universidade Federal do Rio Grande do Norte, 
Natal, RN, 59078-970, Brazil}
\author{M\'at\'e Lencs\'es}
\affiliation
{BME Department of Theoretical Physics,  H-1111 Budapest, Budafoki \'ut 8, Hungary}
\affiliation
{BME ``Momentum'' Statistical Field Theory Research Group, 
H-1111 Budapest, Budafoki \'ut 8, Hungary}
\author{J. C. Xavier}
\affiliation{Universidade Federal de Uberl\^andia, Instituto de F\'isica, C. P. 593, 38400-902 Uberl\^andia, MG, Brazil}
\author{Rodrigo G. Pereira}
\affiliation{International Institute of Physics,  Universidade Federal do Rio Grande do Norte, 
Natal, RN, 59078-970, Brazil}
\affiliation{Departamento de F\'isica Te\'orica
e Experimental, Universidade Federal do Rio Grande do Norte, 
Natal, RN, 59078-970, Brazil}

\begin{abstract} 
 Weakly coupled Ising chains provide a condensed-matter realization of confinement. In these systems,    kinks and antikinks bind into mesons due to an attractive   interaction potential that increases linearly with the distance between the particles. While single mesons have been directly observed in experiments, the role of the multiparticle continuum and bound states  of mesons in the excitation spectrum is far less clear. Using time-dependent density matrix renormalization group methods, we study  the dynamical 
structure factors of one- and two-spin operators in a transverse-field two-leg Ising ladder in the ferromagnetic phase. The propagation of   time-dependent correlations and the   two-spin excitation spectrum  reveal the existence of interchain bound states, which are absent in the one-spin dynamical structure factor. We also identify  two-meson bound states that appear at higher energies, above the thresholds of several two-meson continua. 
\end{abstract}
\maketitle

\section{Introduction}

Over the last few decades, the phenomenon of confinement   has   attracted  considerable interest in both theoretical and experimental   condensed matter physics   
\cite{McCoyWu1978,IsingSpect1,Rutkevich2005,Rutkevich2008,Rutkevich2009,Gabor2017,Collura2019,Verdel2019,James2019,lerose2019quasi,tortora2020,Ishimura1980,Shiba,Lake2010,ColdeaE82010,Morris2014,Moore2011,Grenier2015,Faure2018,Wang2019,Wang2015,Lake2017,
Gannon2019,Liu2019,Lerose2019,Tan2019,fava2020}.  Akin to quark confinement via the  strong force in  quantum chromodynamics (QCD),  the elementary excitations of some quasi-one-dimensional magnets  form bound states  due to interacting potentials that  increase  with the distance between the particles.  As a consequence,  free  particles   cannot be directly  observed in the excitation spectrum. Well established examples include the  ferromagnetic Ising chain in a weak longitudinal field,   the antiferromagnetic XXZ chain with  an easy-axis  staggered field   \cite{Rutkevich2008,Rutkevich2009,Collura2019,Gabor2017, Verdel2019}, and  weakly coupled chains 
\cite{Shiba,Ishimura1980,Lake2010,ColdeaE82010,Morris2014,Wang2015,Grenier2015,Lake2017,Faure2018,Wang2019,Gannon2019,Moore2011}.
In the Ising  chain, for instance, the elementary excitations in the ordered phase are  kinks in the magnetization profile, and   the external longitudinal field creates a linear potential that binds kinks and anti-kinks. 
We shall refer to such bound states as mesons, in analogy with bound states of a quark and an antiquark in QCD. 
Even in the absence of an external magnetic field, a finite  interchain coupling plays the role of the confining potential \cite{Shiba}. 
In addition, confined states arise in the transverse-field  Ising chain with long-range interactions \cite{Liu2019,Tan2019,Verdel2019,Lerose2019} and in 1+1 dimensional quantum electrodynamics \cite{magnifico2019,chanda2019}.

Hallmark signatures of confinement can be observed in dynamical properties of quantum many-body systems in real time as well as in the frequency domain.
In the real-time domain, even a  weak confining potential has been shown to lead to dramatic changes in the quench dynamics, such as non-thermalization, strong suppression of the light cone that bounds the propagation of  correlations, and multifrequency oscillations in the entanglement entropy and one-point functions \cite{Gabor2017,Collura2019,James2019,Verdel2019,lerose2019quasi,tortora2020}. The time evolution after  quantum quenches can be experimentally probed in   quantum simulators, see for example Refs. \cite{Porras2004,Bohnet2016,Marcuzzi2017,Jurcevic2017,Zhang2017}. Recently, the confinement dynamics  was realized in trapped   ions that simulate an Ising-like chain with long-range interactions \cite{Tan2019} and on an IBM quantum computer \cite{Vovrosh2020}. In the frequency domain,  confinement in magnetic systems can be inferred from the 
analysis of dynamical structure factors (DSFs), which relate to  inelastic scattering cross sections   and absorption spectra directly measured in  experiments. From this perspective, the formation of mesons is  manifested as  a discrete spectrum that contrasts with  the two-particle continuum of the   unconfined system~\cite{McCoyWu1978}. Evidence of confinement was   observed in  quasi-one-dimensional  compounds such as CaCu$_2$Co$_3$ \cite{Lake2010}, CoNb$_2$O$_6$ \cite{ColdeaE82010,Morris2014}, BaCo$_2$V$_2$O$_8$  \cite{Grenier2015,Faure2018,Wang2019}, SrCo$_2$V$_2$O$_8$ \cite{Wang2015,Lake2017},  and Yb$_2$Pt$_2$Pb \cite{Gannon2019}. 

In all of these materials, the confinement  of the elementary excitations is an intrinsic property that   arises due to the   interchain coupling. Being weak, the interchain coupling is often treated within a mean-field approximation as an effective magnetic field  proportional to the local magnetization \cite{Shiba}. This approximation leaves out the interesting possibility of multiparticle excitations where bound states   form not only between kinks within the same chain, but also between adjacent  chains. At weak coupling, such interchain bound states can occur in transitions promoted by  two-spin operators that act on different chains simultaneously.  While inelastic neutron scattering is described by one-spin DSFs \cite{Lake2010,ColdeaE82010}, excitations associated with two-spin  operators contribute to the cross section in  resonant inelastic X-ray scattering (RIXS) \cite{Haverkort2010,Ament2011}   and to  the optical conductivity below the Mott gap \cite{Potter2013} measured by   terahertz spectroscopy \cite{Morris2014,Wang2015}.

In this work, we investigate the formation of mesons and multiparticle bound states in weakly coupled transverse-field Ising chains beyond the mean-field approximation for the interchain coupling. For this purpose, we  analyze the DSFs of one- and two-spin operators for a two-leg ladder in the ferromagnetic phase. Since the ladder model is nonintegrable, we  use numerical techniques to compute the physical quantities of interest. In order to determine the ground-state
phase diagram of the system, we apply the density matrix renormalization group (DMRG) \cite{White1992} and truncated fermionic space approach (TFSA) \cite{tfsa}. We   also apply   state-of-the-art time-dependent density matrix renormalization group (tDMRG) methods \cite{FeiguintDMRG}  to compute the dynamical properties. We interpret our results in terms of the semiclassical picture of   massive particles moving in the presence of a linear interaction potential. In addition to the 
mesons observed in the dynamics of ferromagnetic Ising chains, our high-resolution results show that the two-leg Ising ladder harbors additional    bound states  that are not allowed in a single chain  due to  the fermionic  nature of the kinks. Furthermore, we   find peaks   above   two-meson continua which we ascribe to two-meson bound states formed due to a
repulsive meson-meson interaction.

This paper is organized as follows. In Sec. \ref{pd}, we describe the ground-state phase diagram of the two-leg Ising ladder. In addition, we present a discussion of the quantum critical line in terms 
of the scaling field theory. In Sec. \ref{semiclassics}, we discuss the semiclassical problem of two massive 
particles in a linear potential. In Sec. \ref{dsfs}, we present our tDMRG results for three different DSFs that allow us to assess the role of intra- and interchain bound states.  Finally, we provide concluding remarks  in Sec. \ref{conclusions}.

 \section{MODEL AND PHASE DIAGRAM\label{pd}}
 
 The Hamiltonian of the ferromagnetic two-leg  Ising ladder in a transverse magnetic field is  
\be 
H=-J\left[\sum_{j,\alpha}\left(\sigma^x_{\alpha,j}\sigma^x_{\alpha,j+1}+h_z\sigma^z_{\alpha,j}\right)+\lambda\sum_j\sigma^x_{1,j}\sigma^x_{2,j}\right],\label{hamiltonian}
\ee 
where $\sigma^{x,z}_{\alpha,j}$ are   Pauli spin operators acting    on site $j$ of leg $\alpha=1,2$, $J$  is the intrachain exchange coupling,  $h_zJ$ is the transverse 
magnetic field, and $\lambda J$ is the interchain coupling. Throughout this paper, we assume $\lambda,h_z\ge0$ and set the energy scale $J=1$ in the numerical results.

For $\lambda=0$, the system consists of two decoupled transverse field Ising models (TFIMs). The TFIM is  
exactly solvable by mapping, via a Jordan-Wigner transformation,   to free fermions with the dispersion relation \cite{Sachdev2011}
\be 
\varepsilon (k)=2J\sqrt{(h_z-\cos k)^2 + \sin^2 k}.\label{kinkdisprel}
\ee
In this case, the model in Eq. (\ref{hamiltonian})  exhibits a $\mathbb{Z}_2\times \mathbb Z_2$ symmetry corresponding to an invariance under  flipping all the spins in the same chain, $\sigma_{\alpha,j}^x\mapsto -\sigma_{\alpha,j}^x$ $\forall j$.
The   ground-state phase diagram for each decoupled chain  is characterized by two   phases separated by a quantum critical point at $h_z=1$. For $h_z<1$, the system is ferromagnetically ordered and the  ground state has a twofold degeneracy for each decoupled chain. In this case, the   symmetry is   spontaneously broken and the order parameter corresponds to  the magnetization along the longitudinal direction, $\langle \sigma_{\alpha,j}^x\rangle\neq0$. On the other hand, for $h_z>1$ the system is in the paramagnetic phase with unbroken symmetry. A well-known extension of the TFIM involves switching on a longitudinal field $h_x$, in which case the symmetry of the Hamiltonian is explicitly broken and the ground state is unique for any $h_z$. The evolution of the elementary excitations on the $h_z$-$h_x$ plane is   complicated, but there are no additional critical points 
 \cite{tfsa,IsingSpect1,IsingSpect2}.

The situation is quite different in the case of the ladder. Switching on the interchain coupling, $\lambda\neq0$,  lowers the 
symmetry   from $\mathbb{Z}_2\times \mathbb Z_2$ to a single $\mathbb{Z}_2$. The limit   $\lambda \gg 1$ is particularly simple, as the effective Hamiltonian in the low-energy subspace is a single Ising chain with parallel  spins   in each rung.  In fact, the ground-state phase diagram of the two-leg Ising ladder presents a phase
 transition line $\lambda_c(h_z)$ in the Ising universality class, starting from the critical point $\lambda=0$, $h_z=1$. This quantum critical line separates a ferromagnetic from a paramagnetic phase  on the whole $h_z$-$\lambda$ plane \cite{DelMuss2sG1998}.
 Analogously to the TFIM, the remaining  $\mathbb{Z}_2$
  symmetry is spontaneously broken in the ferromagnetic phase and the ground state is twofold degenerate. Meanwhile, the symmetry is preserved in the paramagnetic
  phase with a unique ground state.

\begin{figure}
\begin{centering}
\includegraphics[width=0.45\textwidth]{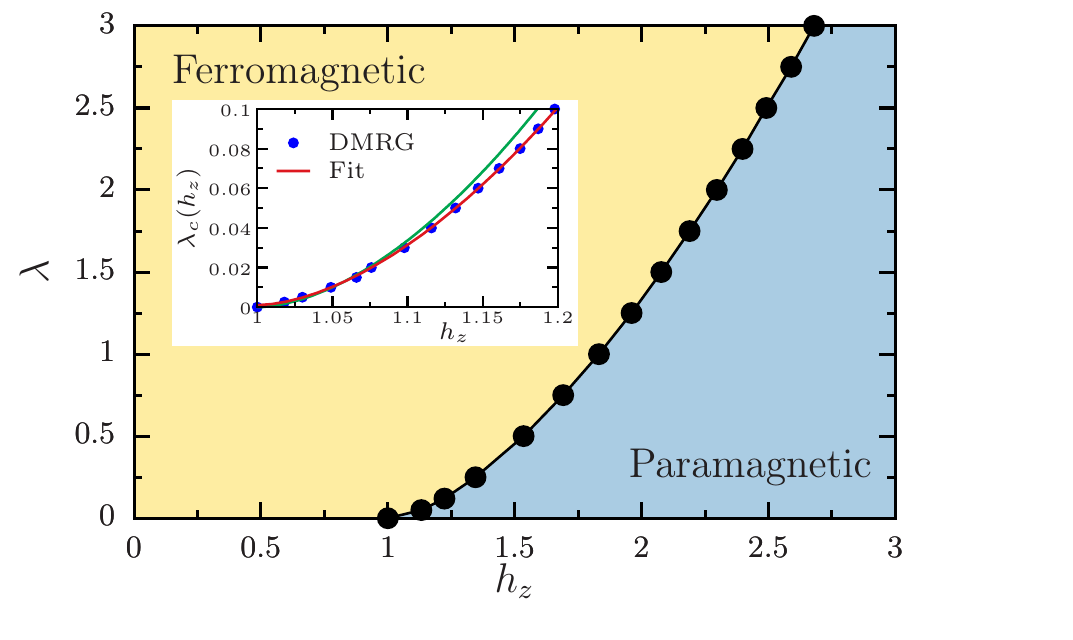}
\caption{Ground-state phase diagram of the two-leg Ising ladder as a function of the
interchain coupling $\lambda$ and the transverse magnetic field $h_z$.
The filled circles are the DMRG points and the black solid line, which connects
those points, is the phase transition line separating the ferromagnetic phase
(in yellow) from the paramagnetic phase (in blue). The inset shows the fit of our numerical results in the interval $0\leq\lambda\leq 0.1$ using   $\lambda_c(h_z)=A(h_z-h_0)^{\beta}$. The estimates of the exponent $\beta$ and the
coefficient $A$ are 1.8 and 1.7, respectively. The solid green line is the field-theory prediction $\lambda_c(h_z)=1.89(h_z-1)^{7/4}$.}\label{phase-diagram}
\par\end{centering}
\end{figure}

The quantum critical line can be accurately determined by analyzing the scaling behavior of the entanglement entropy \cite{Xavier-critpoint}. Following the procedure reported in Ref. 
\cite{Xavier-critpoint}, we   use DMRG to obtain the finite-size estimates $\lambda_c(h_z,L)$, where $L$ is the length of the ladder  with open boundary conditions. In order to obtain $\lambda_c(h_z)$ in the thermodynamic limit, we assume that $\lambda_c(h_z,L)$ behaves as  $\lambda_c(h_z,L)=\lambda_c(h_z)+a/L+b/L^2$. We then estimate $\lambda_c(h_z)$ from the fit of the numerical data considering system sizes $ L= 40, 60, 80, 100$, and $L=200$. The result for the critical line is shown in  Fig. \ref{phase-diagram}. Estimates of critical transverse fields were   obtained for the isotropic (${\lambda=1}$) $N$-leg Ising ladders using the same method \cite{Xavier2014}.

Let us now discuss the model from the scaling field theory point of view. For $\lambda=0$ and $h_z=1$, the system can be described by conformal field theory (CFT) \cite{di1996conformal}.
In this case, the CFT fixed point is a product of two Ising fixed points with scaling fields $\mathbf{1}_\alpha,\epsilon_\alpha,\sigma_\alpha$,  with  $\alpha=1,2$ being  the leg index, corresponding to the identity, energy density and spin density, respectively. The respective
conformal weights $(h,\bar h)$ of these fields are $(0,0)$, $(1/2,1/2)$, and $(1/16,1/16)$. In this context, for $|h_z-1|\ll 1$  and $|\lambda|\ll1$, the ladder can be described by the following Euclidean action
\be
 \label{eq:action}
 S = S_1^{\mathrm{CFT}} + S_2^{\mathrm{CFT}} + \frac{m}{2\pi}\int d^2x (\epsilon_1+\epsilon_2) + \tilde{\lambda} \int d^2x \sigma_1 \sigma_2,
\ee
where $m = 2J(1-h_z)$ is the mass   and  $\tilde{\lambda}=(2/\bar{s}^2)J^{7/4}\lambda$ is the rescaled interchain coupling, where $\bar{s}=2^{1/12}e^{-1/8}\mathcal{A}^{3/2}$ with Glaisher's constant $\mathcal{A}=1.2824271291\dots$ \cite{Wu1976}. For convenience, we define the dimensionless (renormalization-group invariant) combination $\eta\equiv\tilde{\lambda}^{4/7}/|m|$.

\begin{figure}
\begin{centering}
\includegraphics[width=0.45\textwidth]{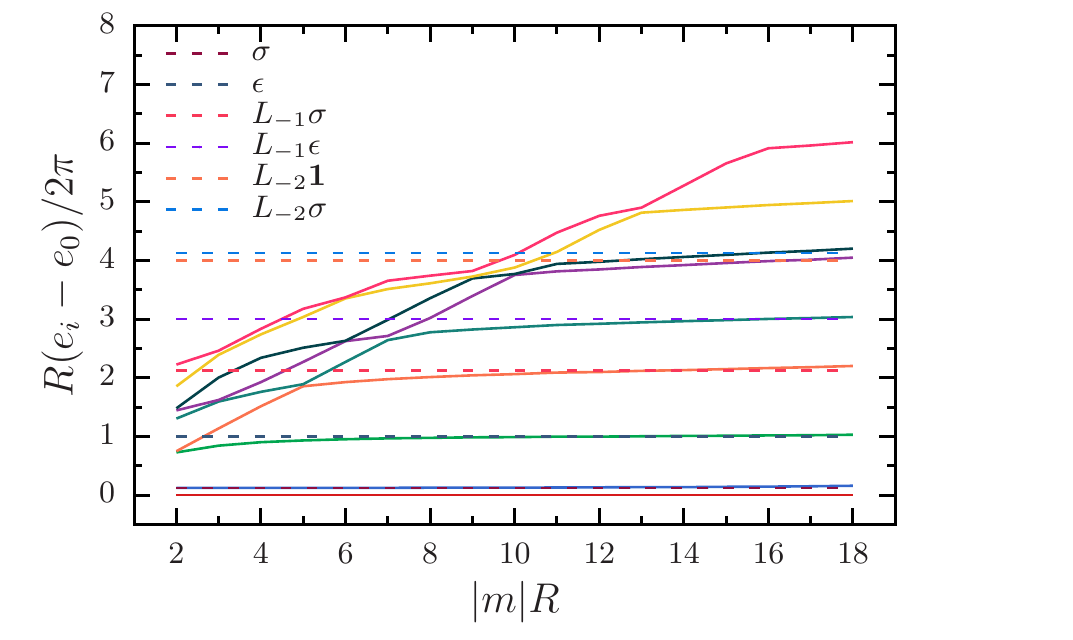}
\caption{The rescaled energy gaps $R(e_i-e_0)/(2\pi)$ at $|m|=1$ and $\tilde{\lambda}=0.61$ against the dimensionless volume $|m|R$ from TFSA.
The dashed horizontal lines show the weights of low-lying Ising CFT states. Here $L_{-n}$ are Virasoro generators. Note that for larger volumes and higher states the TFSA is less reliable
due to truncation errors~\cite{FevGraPeaTotWat2006,GiokasWatts,LencsesTakacs2014,Hogervorst:2014rta}. The lines with slopes are related to the high energy spectrum, and they do not play any role in the phase transition.}
\label{fig:critTFSA}
\par\end{centering}
\end{figure}

The spin-spin coupling in Eq. (\ref{eq:action})  was studied in Ref. \cite{leclair1998}. The phase transition with both the mass term and the spin-spin coupling can be captured in the field theory as  predicted in the context of the two-frequency sine-Gordon model~\cite{DelMuss2sG1998} and later analyzed in Refs.~\cite{FabrGogNer2000,BajPallTakWag2001}.
However, to the best of our knowledge, this has not been studied from the point of view of the coupled Ising field theory. Starting from the paramagnetic phase with a given mass, say $|m|=1$, one hits the 
phase transition line $(\eta=\eta_c)$ by increasing the   interchain coupling. In the paramagnetic phase with $\eta<\eta_c$, the ground state is unique. The   low-energy spectrum of a system with finite volume $R$ exhibits a mass gap with small exponential corrections in $R$.  At the critical point, the energy levels of the finite-volume system   scale with $1/R$.   Finally, for  $\eta>\eta_c$ in the   ferromagnetic phase, the ground state is twofold degenerate up to an exponentially small energy splitting. The evolution of the finite-size spectrum
with increasing $\eta$ can be analyzed using the tensor product extension of the TFSA~\cite{tfsa}. Using TFSA, we have obtained  $\tilde \lambda_c\approx 0.61$ for $|m|=1$. In addition, we are   able to identify the low-lying spectrum 
of the Ising fixed point (see Fig.~\ref{fig:critTFSA}).

Based on the above scaling field theory arguments, one can fit the DMRG data using $\lambda_c(h_z) = A (h_z-h_0)^{\beta} $. The exact critical field at $\lambda=0$ is $h_0=1$. The exact exponent $\beta=7/4$ is fixed by the scaling dimensions of the fields, via the relation with the renormalization-group-invariant parameter $\eta$.  In addition, using the rescaled critical coupling $\tilde \lambda_c\approx 0.62$ for $|m|=1$, we   extract the coefficient $A_{\text{TFSA}} \approx 1.89$ from TFSA data. 
By performing a fit of the DMRG results using the   interval   $\lambda\in[0.025,0.09]$,  we obtain  $\beta_{\text{DMRG}}\approx 1.8$,  $A_{\text{DMRG}}\approx1.7$ and $h_{0}^{\text{DMRG}}\approx0.99$  (see the inset in Fig. \ref{phase-diagram}).  We note  that the fitted exponent varies by about 5$\%$ and the prefactor by about 10$\%$ if we choose different intervals in the range $\lambda\in [0,0.1]$. Since the values of $h_z-1$ used here  are not exceptionally  small, we believe that  more robust and precise fitting parameters could be obtained by incorporating higher-order corrections, associated with irrelevant operators in the field theory, to the expression for $\lambda_c(h_z)$. Note that, in the vicinity  of the Ising critical point, $|h_z-1|\ll 1$, our results show remarkable agreement
with the field-theory prediction.

\section{SEMICLASSICAL SOLUTION\label{semiclassics}}

We now focus on the low-energy excitations of the ferromagnetic two-leg Ising ladder in the    regime of weak interchain coupling. For   comparison with the tDMRG results in Sec. \ref{dsfs}, in the following we   discuss    the confinement effect    based on the mean-field theory picture and the analogy of the Ising ladder with the TFIM in a weak longitudinal  magnetic  field.

Let us recall the basic phenomenology of   confinement in the Ising chain \cite{McCoyWu1978,IsingSpect1,Rutkevich2008}. In the absence of a longitudinal field,  the ferromagnetic Ising chain has two degenerate ground states, 
and the elementary excitations of the system are free domain walls (kinks) interpolating between them. For small but finite $h_x$, the degeneracy is broken and an energy-density difference between
the two ground states is observed, so that one of them becomes a ``false'' one. If we imagine a two-kink configuration on a ``true'' ground-state background, it is clear that it acquires
 additional energy proportional to the distance between the kinks, which then become confined and form mesons.

\begin{figure}
\begin{centering}
\includegraphics[width=0.35\textwidth]{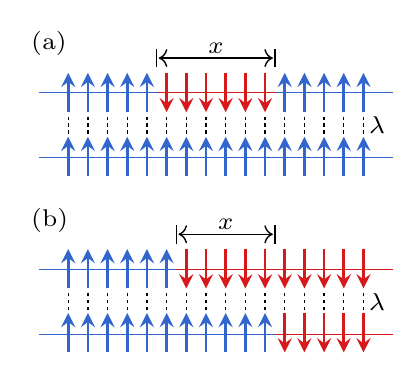}
\caption{Schematic representation of the (a) intrachain and (b) interchain mesons on the  ferromagnetic two-leg Ising ladder. An effective potential proportional to 
the separation $x$ between the kinks  is induced in the region where the spins  are aligned along opposite directions.
}
\label{fig:mesonconf}
\par\end{centering}
\end{figure}

In the case of two weakly coupled chains, one can treat one chain as a source of a longitudinal magnetic field on the other: 
for non-overlapping segments of ground-state configurations, the system is in a ``false''  ground state. Thus, kink confinement can take place in two different ways, as illustrated in  Fig. \ref{fig:mesonconf}.  The   energy cost of the  region with opposite magnetization for spins  on the same rung  is proportional to the   distance between either a kink and an antikink in the same chain [see Fig.  \ref{fig:mesonconf}(a)], or between two kinks in different chains [see Fig.  \ref{fig:mesonconf}(b)].  The former case of an intrachain bound state between  a kink and an antikink is the familiar meson \cite{IsingSpect1}. We shall refer to the latter case as an interchain bound state. Note that, in contrast with mesons, the interchain bound states are topologically charged as they interpolate between two different ground states.

The system  can now be   modeled by a mean-field Hamiltonian in analogy with  the TFIM in a longitudinal field.  
We   write the effective, mean-field two-kink Hamiltonian as \cite{IsingSpect1}
\be
\label{eq:two-kink}
 H_{\text{2k}} = \varepsilon (k_1) + \varepsilon (k_2) + \chi |x_2-x_1|,
\ee
where the dispersions $\varepsilon(k_i)$ are given in Eq.~\eqref{kinkdisprel}, $x_{1,2}$ are the positions of the kinks, and
$\chi = 2\lambda \bra \sigma^x \ket^2$~\footnote{Note that, on a single chain, there is only $\bra \sigma \ket$ in the formula for $\chi$. 
Here, it is squared due to the mean-field approximation.}, with $\bra \sigma^x \ket = (1-h_z^2)^{1/8}$ being the order parameter of the ferromagnetic TFIM \cite{Wu1976}.

The   Hamiltonian in Eq. (\ref{eq:two-kink}) was studied in detail  in Ref.~\cite{Rutkevich2008}. In particular, one can use Bohr-Sommerfeld semiclassical quantization to obtain the  dispersion relations of the bound states. Using the conservation of the total  meson momentum $P = k_1 + k_2$,  the problem is reduced  to solving the      equations \cite{IsingSpect1,Rutkevich2008} 
\be
\label{eq:SCLdisp}
\begin{split}
	2E(P,\nu)k - \int_{-k}^{k} dp\, \Omega(p,P) & = \pi \chi \left(\nu+\frac12\right), \\
	\Omega(k,P) & = E(P,\nu),
\end{split}
\ee
where $\Omega(k,P) = \varepsilon(k+P/2)+\varepsilon(k-P/2)$ and $k\in[0,\pi]$. For  mesons, {\it i.e.}, intrachain bound states, the quantum  number $\nu$ must be an odd integer  in order to obtain  antisymmetric wavefunctions    with respect to exchanging  the positions of the kinks. This constraint stems from the Pauli exclusion principle for kinks in the same chain. We denote the corresponding meson dispersion relations   by $E_n(P)=E(P,\nu=2n-1)$, with $n=1,2,3,\dots$.  By contrast, there is no such constraint for interchain bound states, in which the kinks carry different leg indices, and symmetric wavefunctions are allowed. Thus,  the quantum number $\nu$ can assume any integer values. The dispersion relations for interchain bound states are denoted by $E_n'(P)=E(P,\nu=n-1)$, with $n=1,2,3,\dots$. Note that $E_{2n}'(P)=E_n(P)$, which  means that, within the mean-field approximation of Eq. (\ref{eq:two-kink}),     interchain bound states with antisymmetric wavefunctions are degenerate with   intrachain bound states.

\begin{figure}
\begin{centering}
\includegraphics[width=0.45\textwidth]{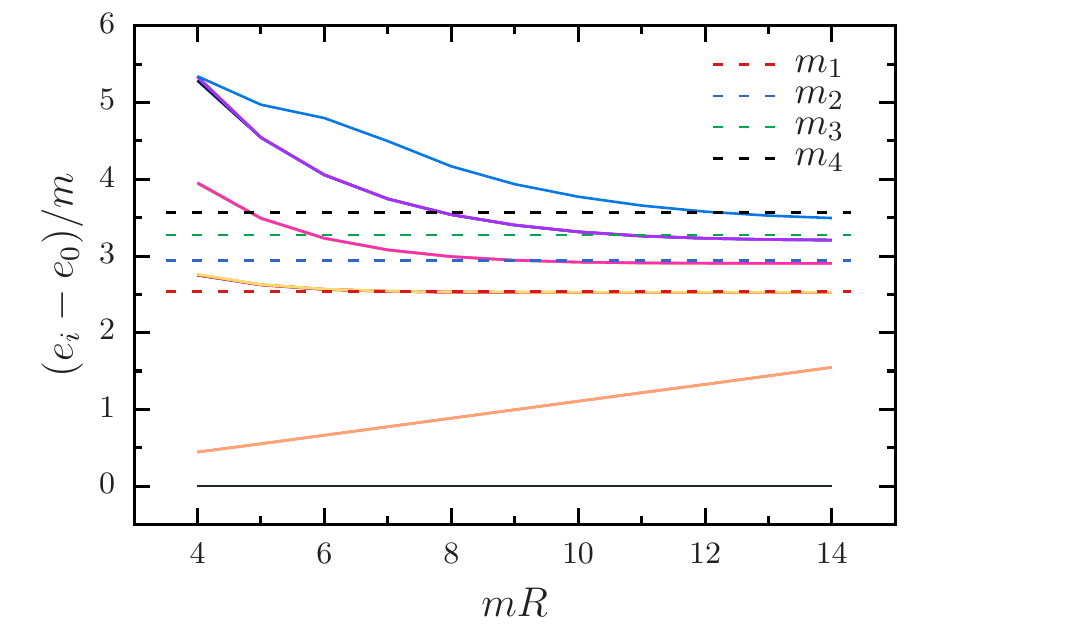}
\caption{ Finite-size energy gaps in the ferromagnetic phase for $\tilde\lambda=0.03$ from TFSA. The ground state is doubly degenerate (horizontal lines at $0$ energy).
The lines with fitted slope $0.111$ correspond to the two degenerate false vacua. Using the analytical energy density $\chi = 2\lambda \bar{s}^2   \approx 0.112$, we also indicate the   meson masses obtained
from the semiclassical approximation $m_n \approx 2m - z_n \chi^{2/3}$.}
\label{fig:mesonTFSA}
\par\end{centering}
\end{figure}

The integral equation in Eq.~\eqref{eq:SCLdisp} can be solved numerically, giving the    dispersion relations $E_n(P)$ and $E'_n(P)$. The minimum of   each dispersion determines  the corresponding particle  masses, which we denote by  $m_n=E_n(P=0)$ for   mesons and $m'_n=E'_n(P=0)$ for interchain bound states. The lightest bound state of all is the first interchain one, with mass $m_1'$.  We have checked the predictions of the semiclassical solution by calculating  the meson masses using  the TFSA. The procedure is the same as for the Ising model in a longitudinal field~\cite{McCoyWu1978,IsingSpect1}. We report the finite size spectrum in the case of $\tilde{\lambda}=0.03$ in the ferromagnetic phase in Fig.~\ref{fig:mesonTFSA}. At  small $\chi$ and for energies near $2m$, as considered in this figure, the   meson masses are well approximated by  $m_n = 2m - z_n \chi^{2/3}$, where $z_n$ is the $n$th zero of the Airy function. Better approximations can be used in the general case, see Refs.~\cite{IsingSpect1,Rutkevich2005,Rutkevich2009,LencsesTakacs2015}.

Some comments are in order. For low energies, Eq.~\eqref{eq:SCLdisp} is only  valid for sufficiently small values of the total momentum. It breaks down when the second derivative  $\partial^2\Omega(k,P)/\partial k^2$ at $k=0$ becomes negative, and the classically allowed regions no longer correspond to the domain of integration used in Eq. (\ref{eq:SCLdisp}) \cite{Rutkevich2008}. Moreover,    for large values of $\nu$  there is no solution to Eq.~\eqref{eq:SCLdisp} because the total momentum cannot be arbitrarily large semiclassically~\cite{Gabor2017}. To circumvent these problems, we   consider solutions of the two-body problem on a lattice described by the     Hamiltonian\bea
H_{\text{2k}}^{\text{lat}}&=&\sum_{j_1,j_2}\sum_{r\geq 0}t_{r} \left(|j_1+r,j_2\rangle\langle j_1,j_2|+|j_1,j_2+r\rangle\langle j_1,j_2|\right.\nonumber\\
&&\left.+|j_1,j_2\rangle\langle j_1+r,j_2|+|j_1,j_2\rangle\langle j_1,j_2+r|\right)\nonumber\\
&&+\sum_{j_1,j_2}\chi |j_2-j_1|\, |j_1,j_2\rangle\langle j_1,j_2|,\label{2bodylattice}
\eea
where $j_1$ and $j_2$ denote the positions of the kinks. The   parameter $\chi$ in the interaction potential is taken from the field theory, as in Eq. (\ref{eq:two-kink}), and a small number of nonzero hopping parameters $t_r$ are fixed so as to approximate   the exact kink dispersion. The eigenstates of the Hamiltonian in Eq. (\ref{2bodylattice}) have wavefunctions of the form $\langle j_1,j_2|\Psi\rangle=e^{iP(j_1+j_2)/2}\phi(j_2-j_1)$, where $P$ is the center-of-mass momentum and $\phi(x)$ is the normalizable wavefunction for the relative coordinate.  Solving  the Schr\"{o}dinger equation for $\phi(x)$ numerically on a finite lattice,  we find that this approach  is in excellent agreement with the semiclassical solution in the momentum range where Eq. (\ref{eq:SCLdisp}) holds, but it  also provides the bound state dispersion relations  in the vicinity of $P=\pi$. We will use these dispersion relations to analyze the tDMRG results  in Sec.~\ref{sec:lam01}.

 \section{DYNAMICAL STRUCTURE FACTORS\label{dsfs}}

We are interested in  the DSFs of one- and two-spin operators for the two-leg Ising ladder described by the Hamiltonian in Eq.  (\ref{hamiltonian}). We 
focus on the parameter regime deep in the ferromagnetic phase, where the elementary excitations can be pictured as domain walls created in pairs by the $\sigma^z$ operator. 
The DSF for  the spin operator on leg $\alpha=1$ is defined as
\be
S^{zz}(q,\omega)=\frac{1}{L}\int_{-\infty}^{\infty}dt e^{i\omega t }\sum_{j,j'}^L e^{-iq(j-j')}C^{zz}(j,j',t),\label{onespindsf}
\ee
with the time-dependent correlations  
\bea
C^{zz}(j,j',t)&=&\bra \Psi_0|\sigma_{1,j}^z(t)\sigma_{1,j'}^z(0)|\Psi_0\ket\nonumber\\
&&-\bra \Psi_0|\sigma_{1,j}^z |\Psi_0\ket\bra \Psi_0|\sigma_{1,j'}^z|\Psi_0\ket.
\eea
Here, $|\Psi_0\ket$ is the ground state and $\sigma_{1,j}^z(t)=e^{iHt}\sigma^z_{1,j}e^{-iHt}$ is the operator $\sigma^z_{1,j}$ evolved in real time. We can write $S^{zz}(q,\omega)$ in the Lehmann representation as
\be
S^{zz}(q,\omega)=\frac{2\pi}{L}\sum_{l>0}|\bra \Psi_l|\sigma^z_{q}|\Psi_0\ket|^2\delta(\omega-E_l+E_{0}),\label{Lehmann1dsf}
\ee
where $\sigma^z_{q}=\sum_j e^{-iqj}\sigma^z_{1,j}$, $|\Psi_l \ket$ are eigenstates of $H$ with energy $E_l$, and $E_{0}$ is the ground-state energy. In the weak coupling regime, $0<\lambda\ll 1$, we expect $S^{zz}(q,\omega)$ to be dominated by excited states in which two kinks created in leg $\alpha=1$ form an intrachain meson with total momentum $P=q$.

 
For two-spin operators, we   consider two distinct DSFs defined as 
\be
S^{4z}_{\text{sl/dl}}(q,\omega)=\frac{1}{L}\int_{-\infty}^{+\infty}dt\,e^{i\omega t}\sum_{j,j'}^L e^{-iq(j-j')}C_\text{sl/dl}^{4z}(j,j',t),\label{twospindsf}
\ee
where \bea
\hspace{-.7cm}C_\text{sl/dl}^{4z}(j,j',t)&=&\langle \Psi_0|O^{\text{sl/dl}}_{j}(t)O^{\text{sl/dl}}_{j'}(0)|\Psi_0\rangle\nonumber\\
\hspace{-.7cm}&&-\langle \Psi_0|O^{\text{sl/dl}}_{j}|\Psi_0\ket\bra \Psi_0|O^{\text{sl/dl}}_{j'}|\Psi_0\ket\label{fourpointfunc}
\eea
are the time-dependent correlation functions for the operators\bea
O^{\text{sl}}_{j}&=&\sigma^z_{1,j}\sigma^z_{1,j+2},\\
O^{\text{dl}}_{j}&=&\sigma^z_{1,j}\sigma^z_{2,j}.\label{corrdl}
\eea
The labels sl and dl in Eqs. (\ref{twospindsf})-(\ref{corrdl}) indicate  that the operators act on two sites located  in the same leg or different legs, respectively. The DSFs  $S^{4z}_{\text{sl/dl}}(q,\omega)$ admit 
 spectral decompositions  analogous to Eq. (\ref{Lehmann1dsf}).  In a simple classical picture of the ferromagnetic phase, the action of $O^{\text{sl}}_{j}$ on the fully polarized state creates four domain walls in the same chain. On the other hand, $O^{\text{dl}}_{j}$ creates two domain walls in each chain. We then expect these DSFs to have significant spectral weight associated with four-kink excitations, which form two mesons or two interchain bound states for $0<\lambda\ll 1$.  Note, however, that two-kink  excitations are also allowed by selection rules. They  are in fact present as contributions in which the four-point function in Eq. (\ref{fourpointfunc}) factorizes into two-point functions, due to the nonzero expectation value of the $\sigma^z$ operator for $h_z\neq0$. 

Since the Ising ladder is nonintegrable, we have used the adaptive tDMRG  \cite{FeiguintDMRG} to compute the DSFs of open ladders. 
While this method is most efficient at treating one-dimensional systems with nearest-neighbor interactions, it is also possible to use tDMRG to investigate narrow ladders. 
To do so, we enlarge the local Hilbert space by combining the rung sites into a supersite, so that the Suzuki-Trotter decomposition can be applied exactly 
in the non-renormalized DMRG sites.

In order to investigate the effects of   a weak interchain coupling on the energy spectrum, we will discuss   the cases of $\lambda=0$ and $\lambda=0.1$. Our
tDMRG results were obtained by setting the ladder length $L=160$ and the transverse magnetic field $h_z=0.5$. The time-dependent correlations were computed by keeping up to 200 
states per DMRG block. The temporal Fourier transforms in Eqs. (\ref{onespindsf}) and (\ref{twospindsf}) were performed in 
the time interval $-t_{\text{max}}<t<t_{\text{max}}$, where $t_{\text{max}}$ is maximum time obtained by tDMRG. To set the maximum time, we have to take into account two limitations. The first one is 
due to the finite length of the ladder. Since we are not interested in boundary effects, $t_{\text{max}}$ is such that the propagation of the correlations does not reach the edges of the chains, 
{\it i.e.}, $t_{\text{max}}<L/(2v)$, where $v$ is the  maximal velocity  that defines the light cone. The second limitation is the numerical errors generated by the truncation procedure and the order of the 
Suzuki-Trotter decomposition. In our computations, the error related to the truncation procedure is smaller than $10^{-7}$ and the time evolution was carried out with second-order Suzuki-Trotter 
decomposition using a  time step $\delta t=0.1$. The maximum time we have considered is in the interval $t_{\text{max}}\in[50,90]$.  We present our numerical results in the following.

\subsection{Decoupled chains}

\begin{figure}
	\begin{centering}
		\includegraphics[width=0.45\textwidth]{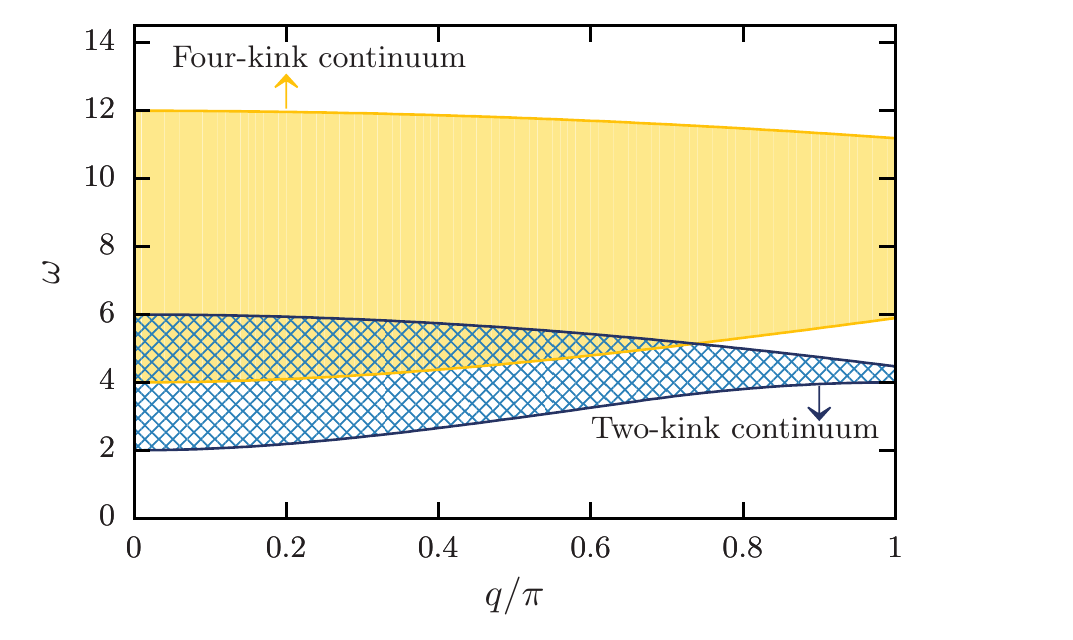}
		\caption{Two- and four-kink excitation continua of the decoupled Ising ladder with $h_z=0.5$. The energies are measured in units of $J=1$.
		The area in blue (yellow) corresponds to the continuum of 
		two (four) kinks. The solid lines indicate the lower and upper thresholds of the continua. 
		\label{kink-cont}}
		\par\end{centering}
\end{figure}

  \begin{figure}
	\begin{centering}
		\includegraphics[width=0.44\textwidth]{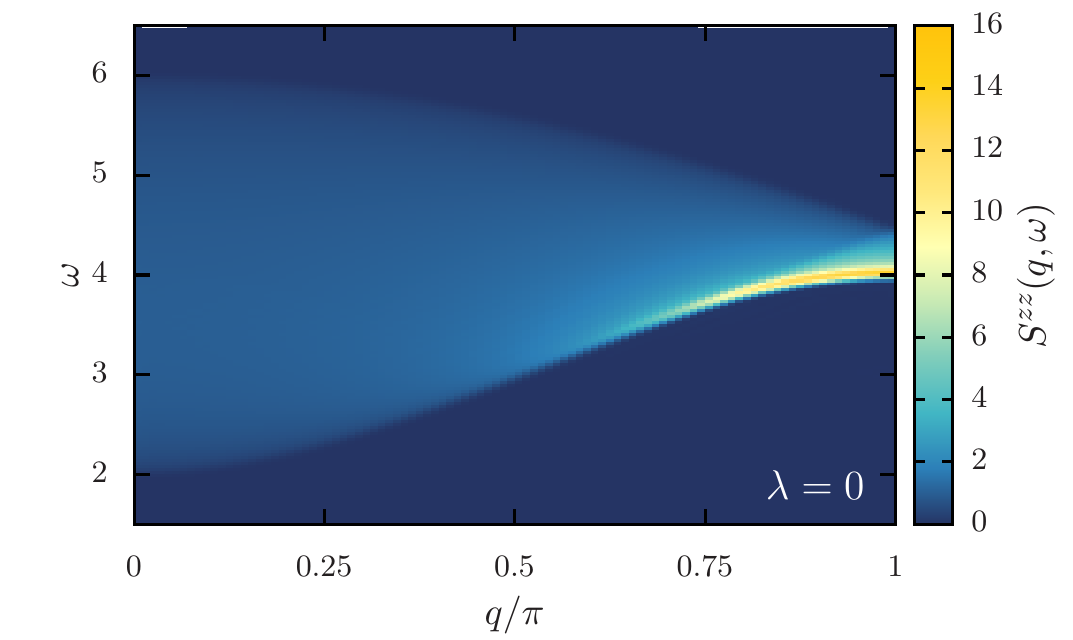}
		\includegraphics[width=0.44\textwidth]{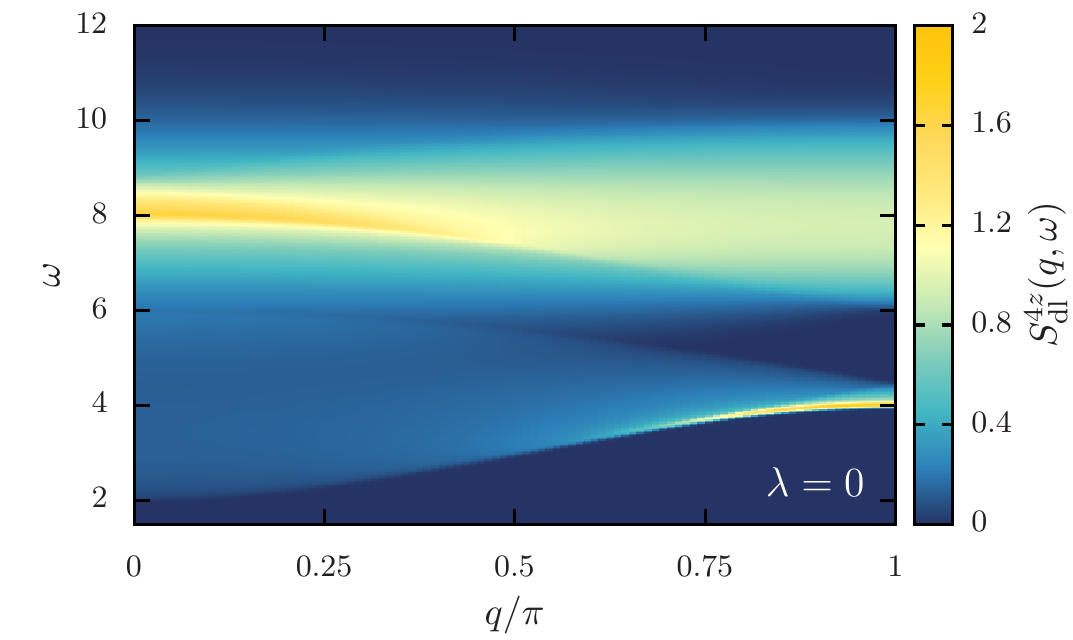}
                \includegraphics[width=0.44\textwidth]{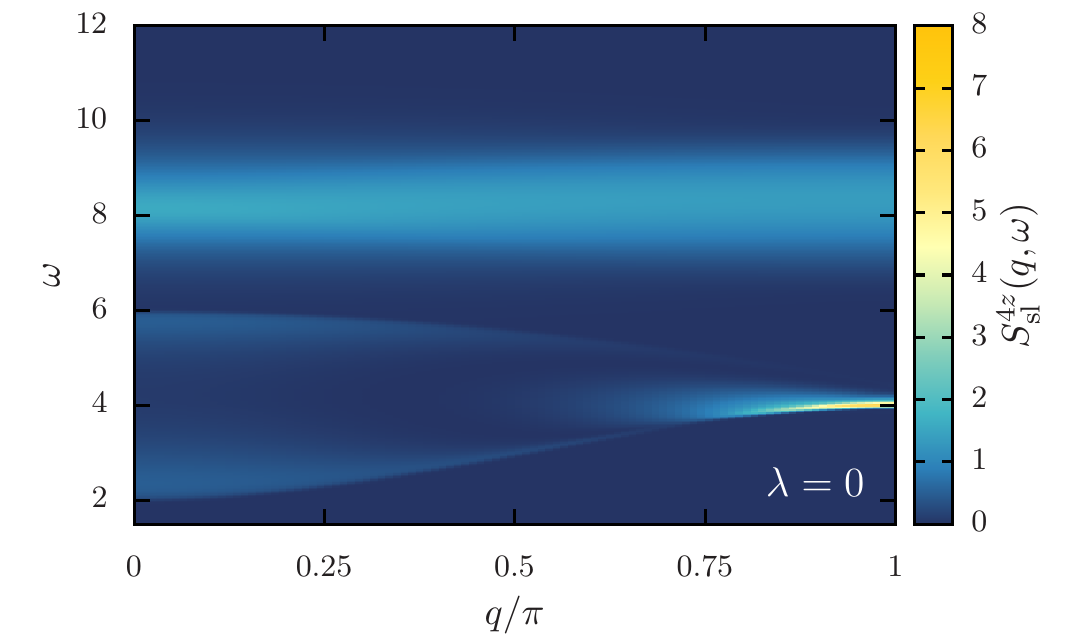}

		\caption {DSFs for  the decoupled two-leg Ising ladder with  $h_z=0.5$, as a function of $q$ and $\omega$. The top panel shows the  one-spin DSF $S^{zz}(q,\omega)$, whose support corresponds to the two-kink continuum. The two-spin DSFs  $S^{4z}_{\text{dl}}(q,\omega)$ (middle) and $S^{4z}_{\text{sl}}(q,\omega)$ (bottom) contain two-kink and four-kink contributions. 
		\label{dsfzzl0}}
		\par\end{centering}
\end{figure}

Let us first discuss  the integrable case $\lambda=0$, corresponding to two decoupled TFIMs. The energy spectrum of the  system can be completely understood from the   dispersion 
relation of free kinks in Eq. (\ref{kinkdisprel}).   In Fig. \ref{kink-cont}, we show the region  in the $(q,\omega)$ plane associated with  two- and four-kink excitations,  which determine   the support of 
$S^{zz}(q,\omega)$ and $S^{4z}_{\text{sl/dl}}(q,\omega)$. Note that the two-kink continuum starts off at $\Delta=2m=4J(1-h_z)$, whereas the four-particle continuum starts at $2\Delta$. The width of the continua depends on the bandwidth of the kink dispersion, which is governed by the transverse field. The upper threshold of the two-kink continuum is given by $\Omega_{2,\text{u}}(q)=2\varepsilon(\pi-q/2)$, and the lower threshold of the four-kink continuum   by $\Omega_{4,\text{l}}(q)=4\varepsilon(q/4)$.  Thus, we  could avoid their overlap  by requiring $\Omega_{4,\text{l}}(0)>\Omega_{2,\text{u}}(0)$, which is verified for $h_z<1/3$. However, a smaller $h_z$ also implies narrower bands for the kinks and slower dynamics.  This requires longer simulation times, making the computation  of the DSFs   unpractical from the tDMRG point of view. Hereafter we shall  consider $h_z=0.5$, as illustrated in Fig. \ref{kink-cont}.  In this case, the two- and four-kink continua overlap  for momentum $q<q_0\approx 0.74\pi$, but they are separated by a gap for $q>q_0$.

\begin{figure}
	\begin{centering}
	\includegraphics[width=0.42\textwidth]{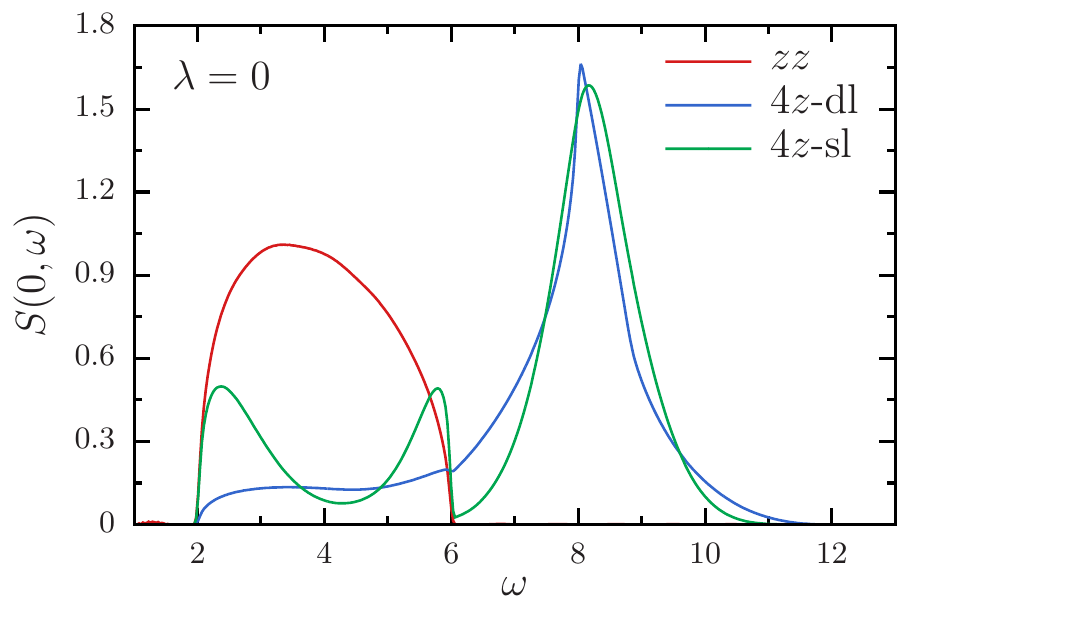}
		\includegraphics[width=0.42\textwidth]{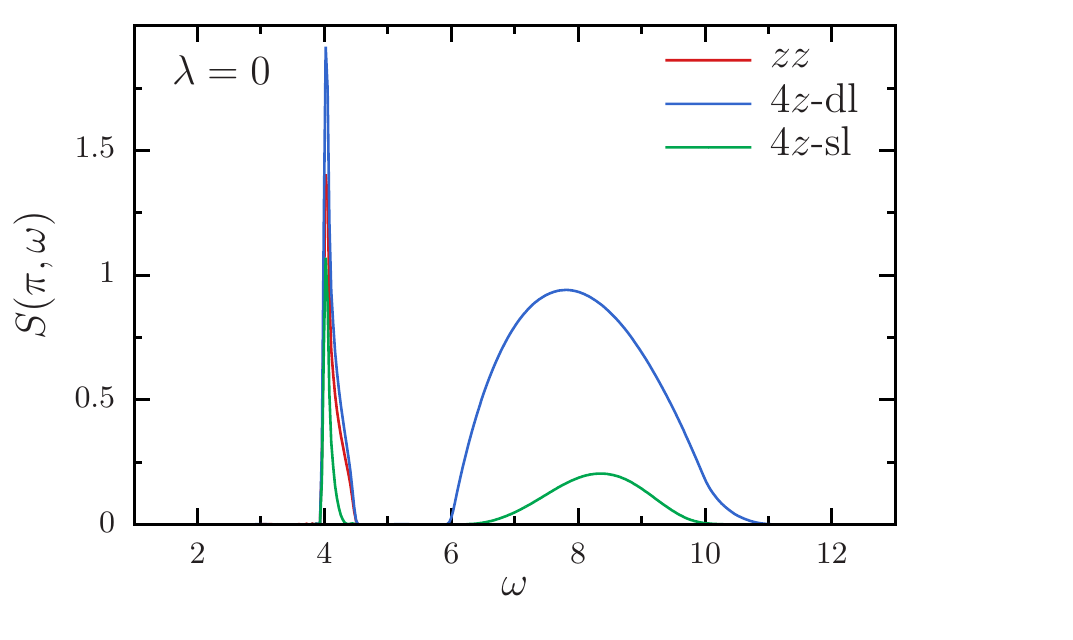}
		
		 \caption {Constant-momentum cuts of the one-spin and two-spin DSFs for $\lambda=0$ and $h_z=0.5$. The top and bottom panels show the line shapes for $q=0$ and $q=\pi$, respectively. The two- and four-kink continua overlap for $q=0$, but they are separated by an energy gap $\Delta_\pi={\Omega_{4,\text{l}}(\pi)-\Omega_{2,\text{u}}(\pi)}\approx1.42$ for $q=\pi$. 
\label{dsf4zl0}}
		\par\end{centering}
\end{figure}

In Fig. \ref{dsfzzl0}, we show the   DSFs  computed by tDMRG for  $\lambda=0$ and $h_z=0.5$. The support of $S^{zz}(q,\omega)$   is precisely the two-kink continuum shown in Fig. \ref{kink-cont}. By contrast, the DSFs $S^{4z}_{\text{sl/dl}}(q,\omega)$ exhibit two- and four-kink contributions (see Figs. \ref{dsfzzl0} and \ref{dsf4zl0}). Despite  the  different spectral weight distributions,  $S^{4z}_{\text{dl}}(q,\omega)$ and  $S^{4z}_{\text{sl}}(q,\omega)$
display the same support, with a significant    weight above the two-kink continuum.  Note the clear separation between the two- and four-kink continua at $q=\pi$.

\subsection{Weakly coupled chains} \label{sec:lam01}

We now consider a  weak interchain coupling $\lambda=0.1$. This coupling introduces an  effective linear potential that  confines the kinks of the Ising chains into bound states. As a result, the two-kink continuum must break up into a series of single-meson peaks.  On the other hand, the four-kink continuum probed by the two-spin DSFs may turn into a  continuum of two propagating bound states, but it may also give  rise to  two-meson  bound states.  Moreover, in the two-leg ladder  we can have both mesons and interchain bound states, and the latter should only be revealed in the DSF $S^{4z}_{\text{dl}}(q,\omega)$. 

\begin{figure}
	\begin{centering}
	\includegraphics[width=0.48\textwidth]{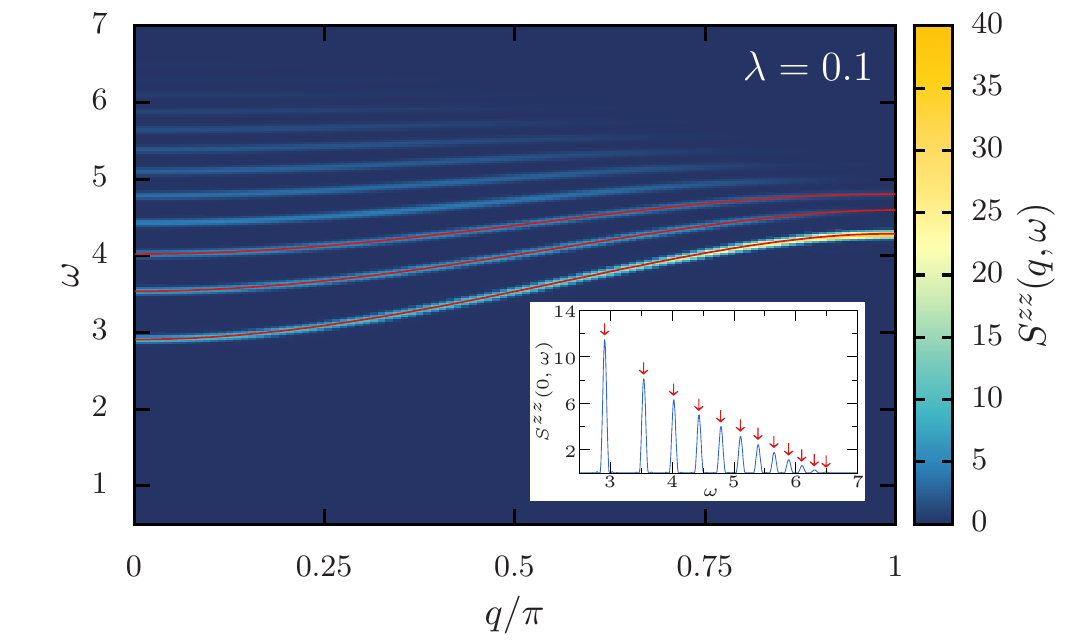}
		\caption {One-spin DSF for $\lambda=0.1$ and $h_z=0.5$, as a function of $q$ and $\omega$.
		The solid red lines represent  the dispersion relations of the intrachain bound states, {\it i.e.}, mesons, extracted from the   solution  of  the two-kink problem on the lattice. 
		The inset shows the line shape of $S^{zz}(q,\omega)$ for $q=0$. The red arrows indicate the masses $m_n$ for the first twelve   mesons.\label{dsfzzl01}}
		\par\end{centering}
\end{figure}

The signatures of the   mesons can be directly  observed in the DSF of the one-spin operator. Indeed, in Fig. \ref{dsfzzl01}  we see a discrete energy spectrum related to the excitations
 created by one spin flip. Our numerical results show a remarkable agreement with the meson dispersion relations calculated as discussed in   Sec. \ref{semiclassics}, even in the regime where the semiclassical approximation  breaks down.  The inset of Fig. \ref{dsfzzl01} shows the line shape of  $S^{zz}(q,\omega)$ for $q=0$, where we compare the peak frequencies with  the predicted    meson masses  $m_n$. Note that the frequency resolution is limited by  the finite time $t_{\text{max}}$ reached by tDMRG, which accounts for the 
  finite width of the meson peaks.

In order to interpret our numerical results for $S^{4z}_{\text{dl}}(q,\omega)$, let us first discuss the case of an open ladder with only one kink in each leg. In this case, the 
  kinks are confined due to the linear potential associated with the region where the chains present opposite local magnetization, as shown in Fig. \ref{fig:mesonconf}(b).  These confined states are interchain bound states. As discussed in Sec. \ref{semiclassics},  now the two-kink wavefunction is not required to be antisymmetric, because the kinks are distinguished by their leg degree of freedom.  Thus, we also take into account the symmetric solutions. An important feature is that the first (lightest) interchain bound state has lower energy  and higher velocity than the lightest meson. Note that single interchain bound states  do not exist in 
ladders with periodic boundary conditions, nor in the bulk   spectrum of open ladders,  because local operators can only create pairs of  kinks     in one or   both 
legs. In particular, the DSF $S^{4z}_{\text{dl}}(q,\omega)$ must contain contributions from excited states with at least two   interchain bound states. However, one can still  detect      interchain bound states  by analyzing the propagation of perturbations in the time-dependent correlations as well as the lower threshold of the two-particle continuum.

\begin{figure}
	\begin{centering}
	\includegraphics[width=0.45\textwidth]{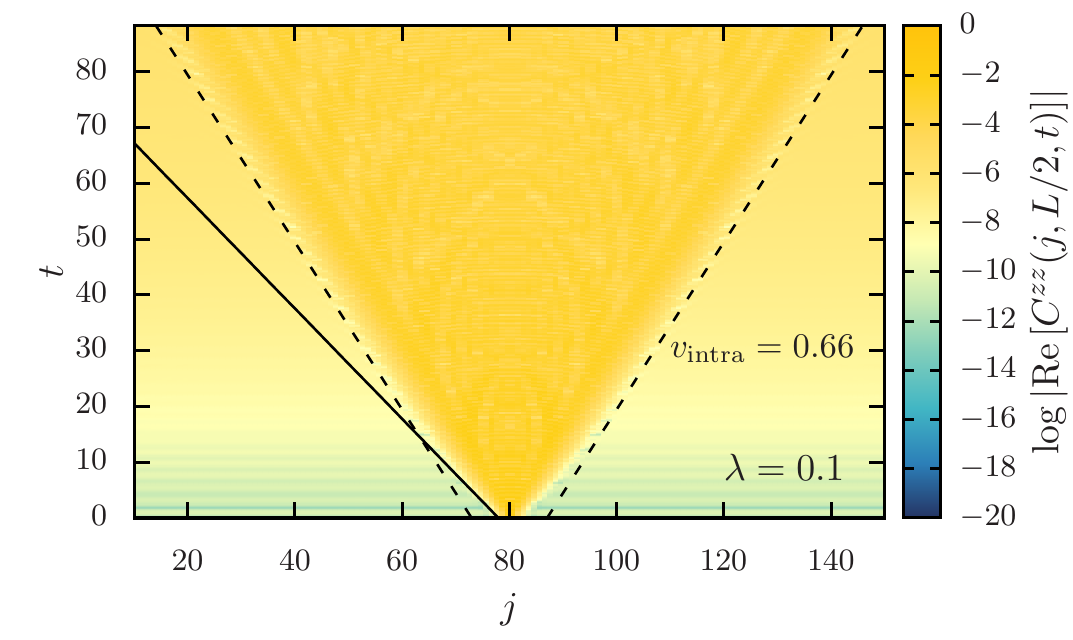}
	\includegraphics[width=0.45\textwidth]{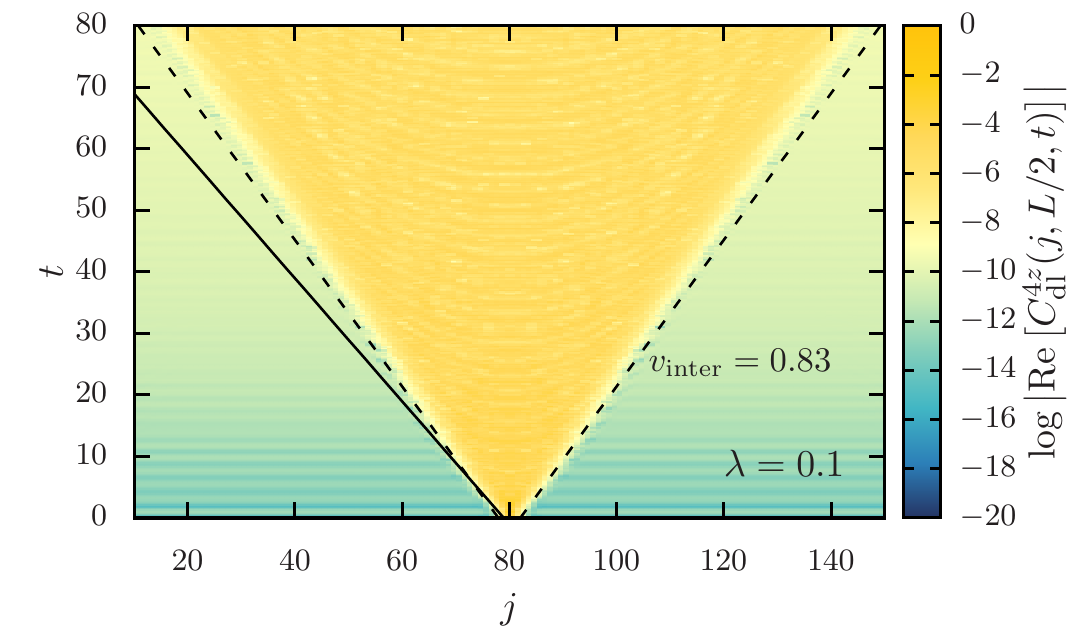}

		\caption {Real part of the time-dependent correlations $C^{zz}(j,L/2,t)$  (top) and $C_\text{dl}^{4z}(j,L/2,t)$ (bottom) as a function of $j$ and $t$. 
		The solids lines are straight lines whose slope is the maximal kink velocity  $v_{\text{kink}}=2h_z=1$. The dashed 
lines are  straight lines with slope $v_\text{intra}$ for  $C^{zz}(j,L/2,t)$ and $v_\text{inter}$ for $C_\text{dl}^{4z}(j,L/2,t)$. The values of  $v_\text{intra}$ and  
$v_\text{inter}$ 
		were acquired from the   bound state dispersion relations.  \label{lightcone}}
		\par\end{centering}
\end{figure}

In Fig. \ref{lightcone}, we show the time-dependent correlations $C^{zz}(j,L/2,t)$ and $C_\text{dl}^{4z}(j,L/2,t)$. To check whether the maximal velocities of the bound states coincide with the ones that bound the light cones, we have computed the   dispersions following the   procedure  described previously. Recall that the dispersion $E_1'(P)$ of the lightest interchain bound state is calculated by taking 
the symmetric ground state wavefunction in the diagonalization of the Hamiltonian in Eq. (\ref{2bodylattice}). For $h_z=0.5$ and $\lambda=0.1$, we find that the maximal intrachain- and 
interchain-bound-state velocities are $v_\text{intra}\approx0.66$ and $v_\text{inter}\approx0.83$. As expected, our results indicate that the fastest particle observed in $C^{zz}(j,L/2,t)$ and 
$C_\text{dl}^{4z}(j,L/2,t)$ are the lightest meson and interchain bound state, respectively.

Another way to identify the  interchain bound state is to look at the continuum of excitations in   $S^{4z}_{\text{dl}}(q,\omega)$. Since the dispersion of  the lightest interchain 
bound state  is below  the intrachain ones, the lower threshold of the  continuum must correspond to scattering states  of two lightest interchain bound states. Similarly to our analysis of   the two-kink continuum in  the decoupled case, we can use the bound state dispersions to construct the  two-particle continua. We label them by the corresponding  masses:   here,  $m'_{n_1}+m'_{n_2}$, with $n_1,n_2\geq 1$,  denotes the continuum defined by one interchain bound state with dispersion $E'_{n_1}(P)$ and another one with dispersion $E'_{n_2}(P)$. Note that there is no  upper threshold in the total two-bound-state continuum, and the separation between different   $m'_{n_1}+m'_{n_2}$ continua is not sharply defined once they overlap and the bound states can decay into lighter particles. Nevertheless, this classification will prove useful in the interpretation of the two-spin DSFs.

\begin{figure}
	\begin{centering}
	\includegraphics[width=0.45\textwidth]{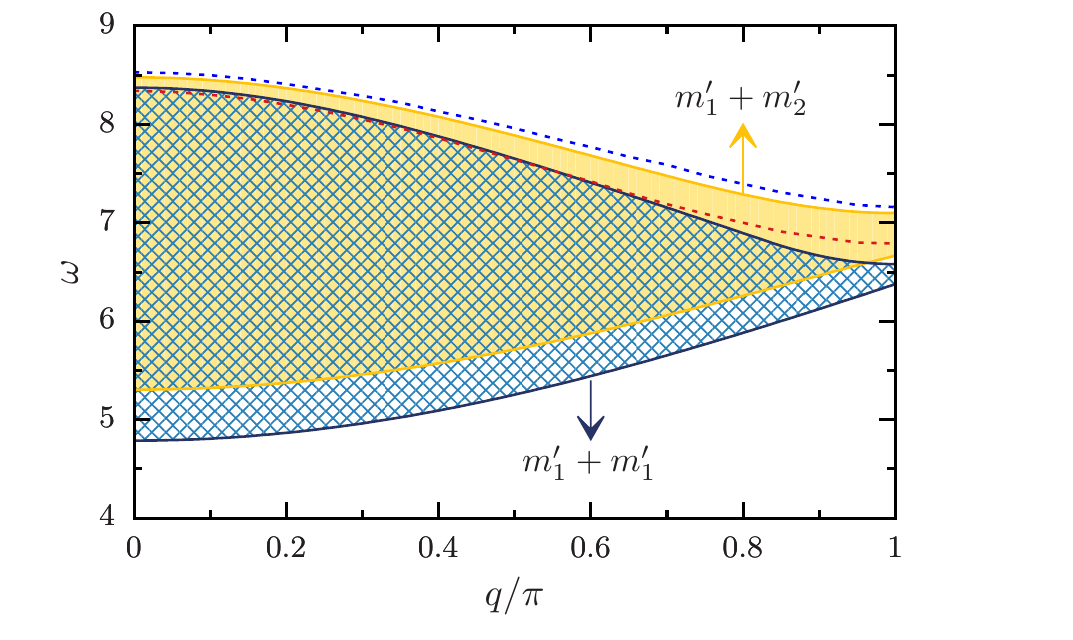}
		\caption { Excitations with two interchain bound states in  the two-leg Ising ladder for $\lambda=0.1$ and $h_z=0.5$, calculated using the bound state dispersion relations. The area in blue  and yellow represent the continua  for  $m'_1+m'_1$ and  $m'_1+m'_2$ excitations, respectively. The dashed lines near the upper thresholds of the continua are  extracted from the bright lines  in the tDMRG results for $S^{4z}_\text{dl}(q,\omega)$ in Fig. \ref{momentumdl4z}.
		 \label{mesoncont}}
		\par\end{centering}
\end{figure}

\begin{figure}
	\begin{centering}
	\includegraphics[width=0.45\textwidth]{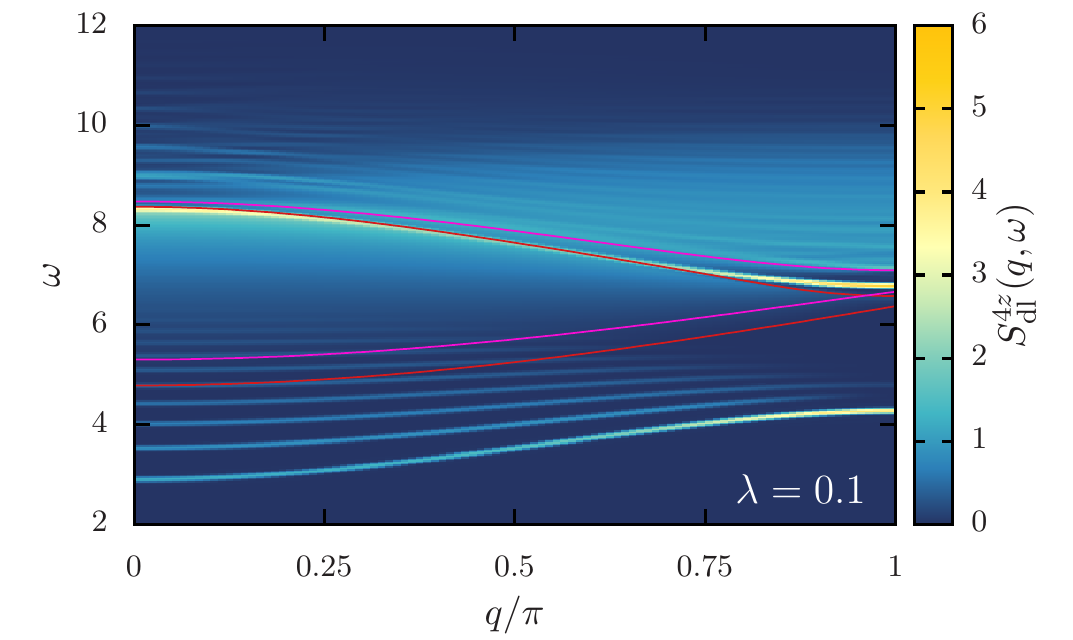}
	\includegraphics[width=0.45\textwidth]{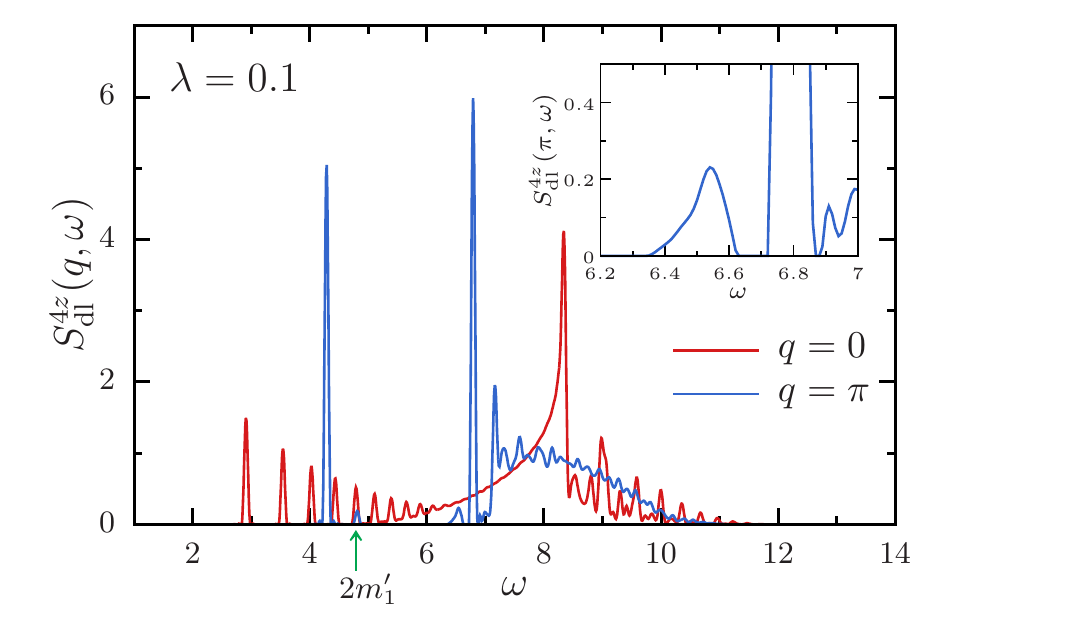}
		\caption {(Top) Two-spin DSF $S^{4z}_\text{dl}(q,\omega)$ for $\lambda=0.1$ and $h_z=0.5$,  as a function of $q$ and $\omega$. The solid red lines are the    lower and upper thresholds of the 
two-interchain-bound-state continuum $m'_1+m'_1$. 
		The pink   lines are the     thresholds of the continuum $m'_1+m'_2$. (Bottom) Constant-momentum cuts of the DSF at $q=0$ and $q=\pi$. 
		The green arrow indicates the lower threshold of the continuum for $q=0$, given by   twice the mass of the lightest interchain
		bound state. For $q=\pi$, there is a pronounced peak    above the  $m'_1+m'_1$
		continuum. 
		The inset is a zoom in of a frequency interval around this peak. The   predictions for the lower and upper threshold of the  $m'_1+m'_1$ continuum at $q=\pi$ are
		$\omega \approx 6.38$ and $\omega \approx 6.58$, respectively. 
		 \label{momentumdl4z}}
		\par\end{centering}
\end{figure}

The $m_1'+m_1'$ and $m_1'+m_2'$ continua are represented  in Fig. \ref{mesoncont}. The lowest excitation occurs at  the  frequency    
$\omega=2m_1'$, corresponding to the energy cost of creating  two   interchain bound states with momentum $q=0$.  For $\lambda=0.1$ and $h_z=0.5$, we find  $m_1'=E_1'(0)\approx 2.39$.  
 This mass is consistent with the lower threshold of the continuum present in the DSF $S^{4z}_{\text{dl}}(q,\omega)$ calculated by tDMRG (see   Fig. \ref{momentumdl4z}).  At $q=0$, the peaks that appear  below and slightly above $2m_1'$ can be identified with  single-meson excitations,  which descend from the two-kink continuum and have   the same frequencies observed  in $S^{zz}(q,\omega)$.  Remarkably, for all values of $q$ the spectral weight associated with  the  excitation of two bound states vanishes as the frequency approaches the lower threshold of the $m_1'+m_1'$ continuum, and there is no evidence for a  bound state of two interchain bound states below the continuum.

\begin{figure}
	\begin{centering}
	\includegraphics[width=0.45\textwidth]{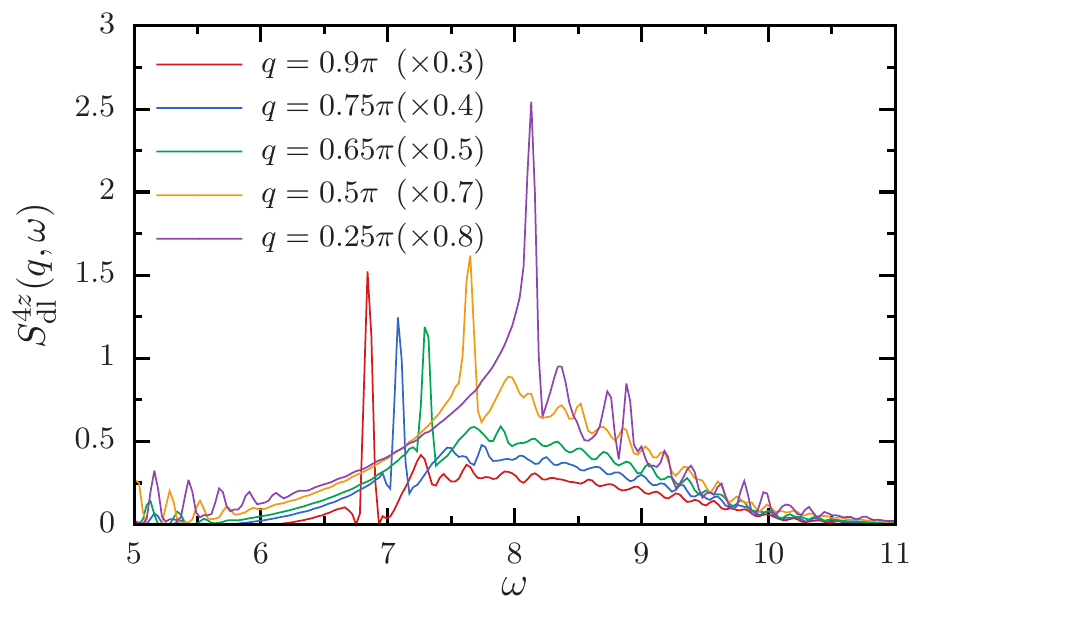}
		\caption { Line shape of  $S^{4z}_\text{dl}(q,\omega)$ for $\lambda=0.1$, $h_z=0.5$, and five different values of $q$.  As $q$ increases, a narrow peak emerges  from the  $m_1'+m_1'$ continuum below it. 
		 \label{dl4zl01}}
		\par\end{centering}
\end{figure}

A prominent feature in the line shape of the DSF $S^{4z}_{\text{dl}}(q,\omega)$ for ${q=0}$, see the bottom panel in Fig. \ref{momentumdl4z}, is   the steep rise in the   intensity     as the frequency increases above $2m_1'$, terminating in a highly asymmetric peak near  the upper threshold of the $m_1'+m_1'$ continuum. By contrast, for $q=\pi$ the $m_1'+m_1'$ continuum covers only a narrow frequency range, in agreement with the result in Fig. \ref{mesoncont}, and has a rather small spectral weight. However, we observe a sharp peak \emph{above} the $m_1'+m_1'$ continuum. The amplitude of this peak is even higher than that of the first   meson peak in $S^{4z}_{\text{dl}}(q=\pi,\omega)$. To track the evolution of the new peak, in Fig. \ref{dl4zl01} we show the line shapes of  $S^{4z}_\text{dl}(q,\omega)$  for different values of momentum.  As one moves from $q=0$ to $q=\pi$, the   peak appears to split off from the continuum below it.  For this reason, and because the width is within the frequency resolution of the tDMRG simulations, we interpret this peak as a  bound state of two interchain bound states. 

On general grounds, we expect   mesons  to interact via a short-range potential, since, unlike the linear potential responsible  for kink confinement, the interaction energy does not increase with the distance between mesons. The same is true for interchain bound states in the two-leg ladder.  The presence of a bound state of two interchain bound states, or rather an \emph{antibound} state,  above the $m_1'+m_1'$ continuum suggests that their  interaction is predominantly  repulsive. The effective scattering amplitude in the Bethe-Salpeter equation that describes the bound state formation must  be a function of the  momenta of the interchain bound states, and  also  depend on  the internal wavefunction $\phi(x)$ for the   relative coordinates between kinks, but we do not attempt to calculate it here. It is important to note that the formation of   bound states of bound states should not be restricted to the $m_1'+m_1'$ continuum, but may occur for all   $m'_{n_1}+m'_{n_2}$ excitations. Thus, one possible explanation for the irregular line shape of  $S^{4z}_{\text{dl}}(q,\omega)$ above the $m_1'+m_1'$ continuum is that it stems from  a series of overlapping two-particle continua and the associated   bound states with finite lifetimes. This is consistent with a second peak found just above the $m_1'+m_2'$ continuum, see Figs. \ref{mesoncont} and \ref{momentumdl4z}.

\begin{figure}
	\begin{centering}
	\includegraphics[width=0.45\textwidth]{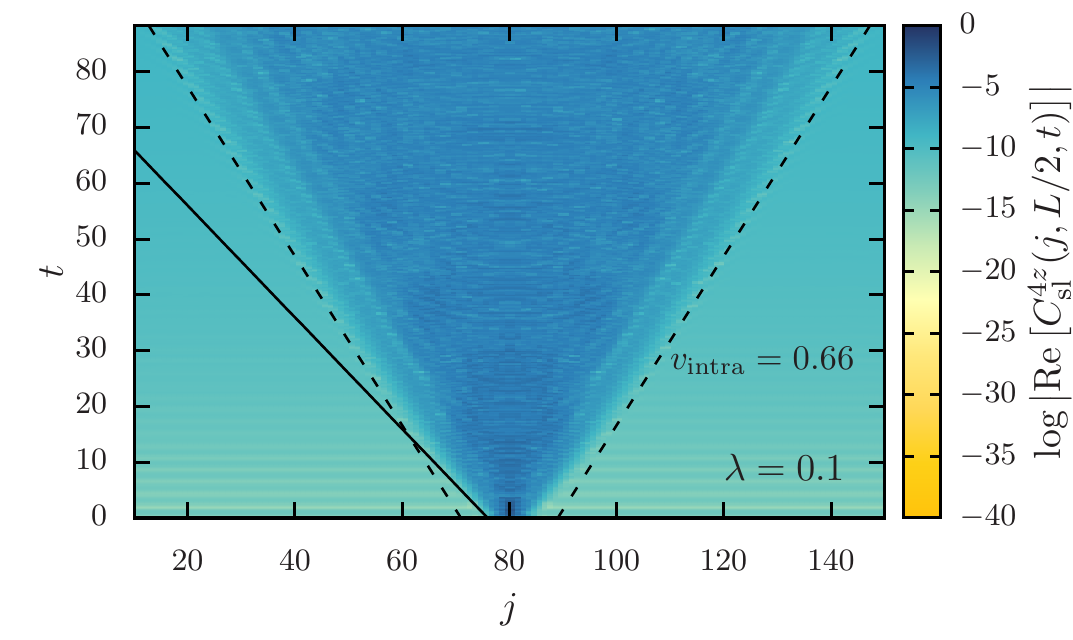}
		\caption {Real part of the time-dependent correlation $C_\text{sl}^{4z}(j,L/2,t)$ as a function of $j$ and $t$.
		The solid and dashed lines are straight lines with slope $v_{\text{kink}}=1$ and $v_{\text{intra}}=0.66$, respectively. \label{lightconesl}}
		\par\end{centering}
\end{figure}

Finally, let us discuss the tDMRG results for $S^{4z}_{\text{sl}}(q,\omega)$. In contrast to $S^{4z}_{\text{dl}}(q,\omega)$,  the spectrum of  $S^{4z}_{\text{sl}}(q,\omega)$ does not contain interchain bound states  because the two spin 
flips are performed  on the same leg. For this reason, only the antisymmetric wavefunctions are allowed in the solutions of  Eqs. (\ref{eq:SCLdisp}) and (\ref{2bodylattice}). Indeed, from the propagation of the time-dependent 
correlation shown in  Fig. \ref{lightconesl}, we   confirm that the fastest particle in this case  is the lightest  intrachain bound state with  maximal velocity    $v_\text{intra}\approx 0.66$.  

\begin{figure}
	\begin{centering}
	        \includegraphics[width=0.45\textwidth]{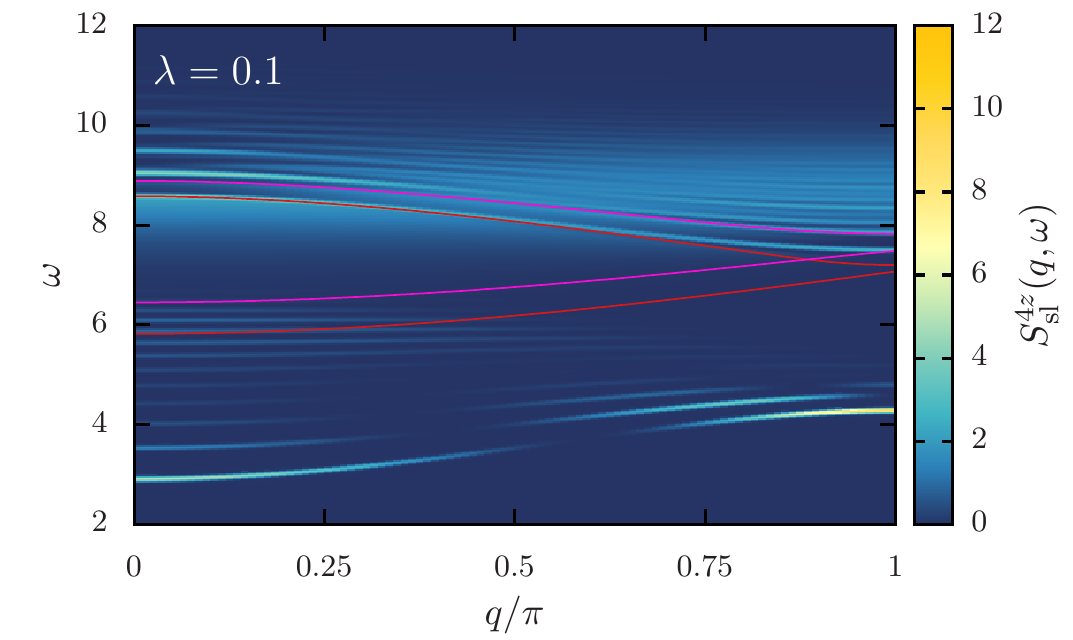}
		\includegraphics[width=0.45\textwidth]{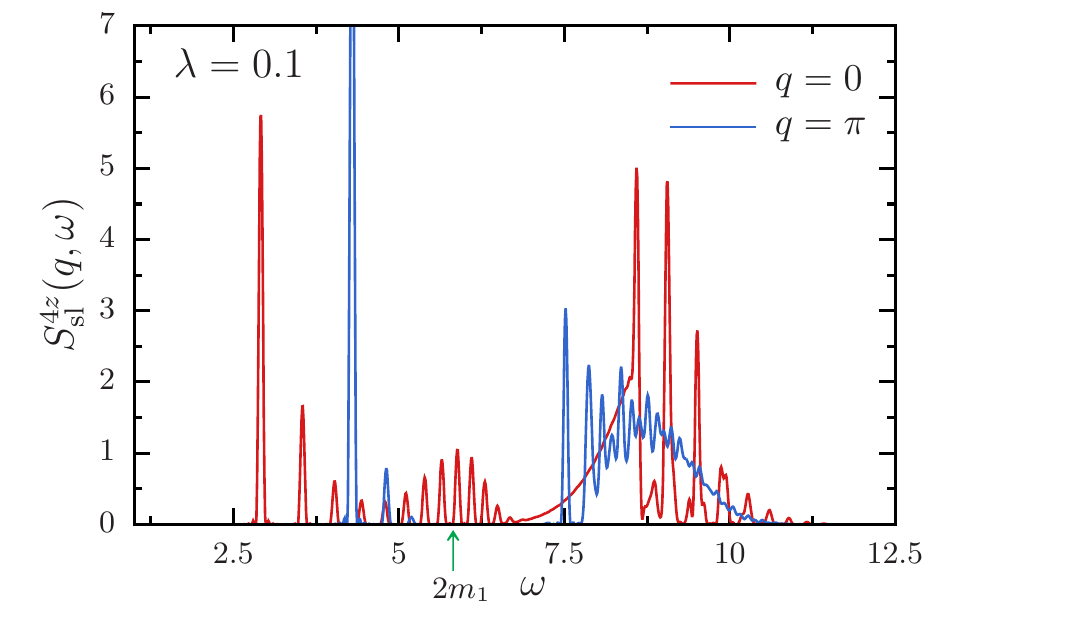}
		\caption {(Top) Two-spin DSF $S^{4z}_\text{sl}(q,\omega)$ as a function of $q$ and $\omega$. The solid red lines are the thresholds of the $m_1+m_1$ continuum. The pink  lines are the     thresholds of the $m_1+m_2$  continuum. (Bottom) Line shapes of  $S^{4z}_\text{sl}(q,\omega)$ for   $q=0$ and $q=\pi$. The green arrow at $\omega=2m_1$ marks the lower threshold of the two-meson continuum for $q=0$. \label{sl4zl01}}
		\par\end{centering}
\end{figure}

Figure \ref{sl4zl01} shows the results for  $S^{4z}_{\text{sl}}(q,\omega)$. The first remark is that this DSF also exhibits single-meson peaks at the same frequencies as the  one-spin DSF. However, in this case the amplitude of the meson peaks for $q=0$ is a nonmonotonic function of frequency, behaving similarly to the two-kink contribution in $S^{4z}_{\text{sl}}(q=0,\omega)$ for the decoupled ladder (see Fig. \ref{dsf4zl0}).  In addition,  the lower threshold  of  the two-meson continuum at $\omega=2m_1$ is not apparent in the line shape for $q=0$. The reason is  that the spectral weight vanishes more rapidly than in the case of $S^{4z}_{\text{dl}}(q,\omega)$ as the frequency approaches the lower threshold of the $m_1+m_1$ continuum. This was already noticeable in the DSF for decoupled chains in Fig. \ref{dsf4zl0}, and follows from the stronger effect of the Pauli exclusion principle when all four kinks propagate in the same leg.  Also in Fig. 
\ref{sl4zl01}, we show the line shape of  $S^{4z}_{\text{sl}}(q,\omega)$  for  $q=\pi$. Again, we observe a series of peaks above the upper thresholds of the $m_{n_1}+m_{n_2}$ continua, and attribute them to bound states of two   mesons with finite lifetimes.  The multiple peaks can be discerned more clearly in this  case as they are more widely separated than in $S^{4z}_{\text{dl}}(q,\omega)$, since here we have only half as many  continua of two propagating bound states. 

\section{Conclusions\label{conclusions}}
We  have studied the formation of intra- and interchain bound states in the  ferromagnetic phase of the transverse-field two-leg Ising ladder. First, we mapped out the phase diagram of the model using DMRG and analyzed the critical line using the scaling field theory and the truncated fermionic space approach. We then used the adaptive tDMRG method to calculate  three different dynamical structure factors.  The one-spin dynamical structure factor probes excitations with two domain walls, or kinks, in the same leg of the two-leg Ising ladder. The other two dynamical structure factors, defined for two-spin operators, probe excitations with four kinks,   either all    in the same leg or two pairs  in different legs. 

The spectrum of the one-spin  operator is completely described by the confinement of  kinks and anti-kinks into single   mesons. A new family of excitations, which we call interchain bound states, appears  in the spectrum of the two-spin operator acting on both legs.  The lightest interchain bound state has a smaller  mass  than the lightest meson because its wavefunction is not required to vanish when the two kinks are in the same rung, but in different legs. As a consequence,  the interchain bound states  determine the lower threshold of the two-particle continuum. The two-spin dynamical structure factors also exhibit a series of higher-energy peaks associated with   bound states of bound states, which are two-meson bound states in the case where spin operators act on sites within the same chain. 

Our results should be relevant to the interpretation of experiments on weakly coupled Ising chains. In fact, the existence of a hierarchy of bound states was proposed based on the experiments  reported in  Ref. \cite{Morris2014}. It would be interesting to quantify the contribution from different two-spin operators to  the   absorption spectra measured by   terahertz spectroscopy  and investigate the role of the two-meson continuum and of the interchain bound states. 

Finally, we would like to briefly comment on the generalization of our results  to $N$-leg ladders and three-dimensional arrays of weakly coupled chains, as found in real materials. The interchain bound state can be generalized to bound states of $N$ kinks, but creating them with a significant amplitude  requires    an operator that  acts simultaneously on $N$ spins in different chains. The resulting excitation can be interpreted as a domain wall whose energy increases with $N$. On the other  hand,   we may consider  the fate of  interchain bound states composed of two kinks or two antikinks when we increase the number of coupled chains. In this case, the neighboring chains which are not excited introduce a linear potential that  confines pairs of interchain bound states with opposite topological charge. As a result, while the two-meson continuum survives in the presence of  more chains, we  expect   the continuum of two  interchain bound states to be completely replaced by bound states of bound states.

{\it Note added}: Recently, we became aware of a related work by G. Lagnese {\it et al.} \cite{Lagnese}, who independently developed the idea of intrachain and interchain bound states in a different ladder system.

\begin{acknowledgements}	
We would like to thank  A. Tsvelik for proposing and discussing   a related problem, and  G. Tak\'acs, D. X. Horv\'ath and G. Z. Feh\'er for discussions on the truncation method.  We are also grateful to S. Rutkevich for comments on the terminology. We acknowledge financial support from  the Brazilian agencies  FAPEMIG (J.C.X.) and 
CNPq (R.G.P.).  We thank the High Performance Computing Center (NPAD) at UFRN for providing computational resources. Part of M.L.'s work was carried out at the International Institute of Physics, 
which  is supported by the Brazilian ministries MEC and MCTIC. M.L. also acknowledges support provided from the National Research,
Development and Innovation Office of Hungary, Project no. 132118 financed under the PD\_19 funding scheme.
\end{acknowledgements}

\bibliographystyle{apsrev4-1}
\bibliography{isinglad}

\end{document}